# Delay Analysis of Graphene Field-Effect Transistors

Han Wang, Allen Hsu, Dong Seup Lee, Ki Kang Kim, Jing Kong and Tomas Palacios

*Abstract—* In this letter, we analyze the carrier transit delay in graphene field-effect transistors (GFETs). GFETs are fabricated at the wafer-scale on sapphire substrate. For a device with a gate length of 210 nm, a current gain cut-off frequency $f_T$ of 18 GHz and 22 GHz is obtained before and after de-embedding. The extraction of the internal ($C_{gs,i}$, $C_{gd,i}$) and external capacitances ($C_{gs,ex}$ and $C_{gd,ex}$) from the scaling behavior of the gate capacitances $C_{gs}$ and $C_{gd}$ allows the intrinsic ($\tau_{int}$), extrinsic ($\tau_{ext}$) and parasitic delays ($\tau_{par}$) to be obtained. In addition, the extraction of the intrinsic delay provides a new way to directly estimate carrier velocity from the experimental data while the breakdown of the total delay into intrinsic, extrinsic, and parasitic components can offer valuable information for optimizing RF GFETs structures.

## I. INTRODUCTION

The extraordinary transport properties of graphene [1], together with its excellent chemical and mechanical stability, have motivated the development of graphene-based radio frequency (RF) electronics [2][3]. Recent work has demonstrated graphene field-effect transistors (GFETs) with current-gain cut-off frequency ($f_T$) in the hundreds of gigahertz range [4][5][6].

Besides extracting figures of merits such as $f_T$ and maximum oscillation frequency ($f_{max}$), the high frequency performance of the GFETs can also be investigated by extracting its carrier transit delays and by understanding how the delay depends on the intrinsic and extrinsic properties of the device. Such analysis not only gives deep physical insight into the carrier transport in the channel, but also provides valuable information that can guide the device engineers in designing high performance RF GFETs. The contribution of this letter is three-fold. First, GFETs are fabricated on sapphire substrate [7] to reduce the parasitics from the ground-signal-ground (GSG) probe pads. This minimizes the error in measuring the S-parameter of the device and allows small-signal capacitances to be accurately extracted. Second, we present for the first time a detailed delay analysis of high frequency graphene transistors. Lastly, the simple and robust method proposed can accurately extract the intrinsic transit delay of the GFETs - the delay purely associated with the carrier transiting across the intrinsic gate region – and allows a new method for direct experimental extraction of the average carrier velocity in the channel. In addition, the individual contributions from the intrinsic, extrinsic and the parasitic elements to the total carrier transit delay can be estimated, which provides valuable information for optimizing the design of RF transistors.

## II. GRAPHENE FETS ON SAPPHIRE SUBSTRATE

The delay analysis relies on accurate two-port S-parameter measurements of the transistor, from which the small-signal capacitances between various electrodes of the active device can be extracted using small-signal equivalent circuit models. Most of today's graphene devices are fabricated on thermally grown $SiO_2$ on Si substrates for two reasons: (1) the ability to identify single- and few-layer graphene sheets using a standard optical microscope and (2) the ability to use the conductive silicon as a back-gate. Unfortunately, for GFETs on a conductive substrate, such as doped silicon, the active device is embedded in the large parasitics of its GSG probe pads. The de-embedding process, hence, involves subtraction between two large numbers, which can lead to significant errors in the de-embedded S-parameters and a large ratio between these devices' $f_T$ values before and after de-embedding [5]. These errors in the de-embedded S-parameters will be carried over to the extraction of the capacitances and make delay analysis virtually impossible.

To reduce the GSG probe pad capacitances and improve the accuracy of the de-embedded S-parameter, we fabricate RF GFETs on a sapphire wafer (500 μm thick) with substrate resistivity above $10^{16}$ Ω.cm. For comparison, the resistivity of conductive Si is less than 1 Ω.cm and about $10^3$ Ω.cm in high-resistivity Si. The highly resistive sapphire substrate can help eliminate most of the capacitances contributed by the coupling between the pad metals and the charge carriers in the substrate.

The graphene used in this work is grown by chemical vapor deposition (CVD) method on copper catalyst [8]. Films are then transferred to a sapphire substrate [8]. Single-layer graphene has been obtained, which uniformly covers more than 95% area of the sample. Figure 1(a) shows the Raman spectrum of single-layer graphene on sapphire substrate taken with a 532 nm excitation laser. The Raman spectrum of graphene-on-sapphire is almost identical to the Raman spectrum of graphene-on-$SiO_2$ except for a broad background fluorescence commonly found in sapphire due to trace impurities [9]. Room-temperature carrier mobilities were in the range of 2,234±95 $cm^2$/V.s (for a sheet charge density $n_S$= 6.0±0.4×$10^{12}$ $cm^{-2}$) as measured through van der Pauw structures (B=0.3 T, I=0.1 mA). As reference, mobilities in graphene transferred on to 300 nm silicon dioxide are typically 2,220±174 $cm^2$/V.s (for a sheet charge density $n_S$= 5.5±0.6×$10^{12}$ $cm^{-2}$). To fabricate graphene transistors, the ohmic contacts of the GFETs are first formed by depositing a 2.5 nm Ti/45 nm Pd/15 nm Au metal stack by e-beam evaporation using a pre-ohmic aluminum capping process [10]. Device isolation is achieved by $O_2$ plasma

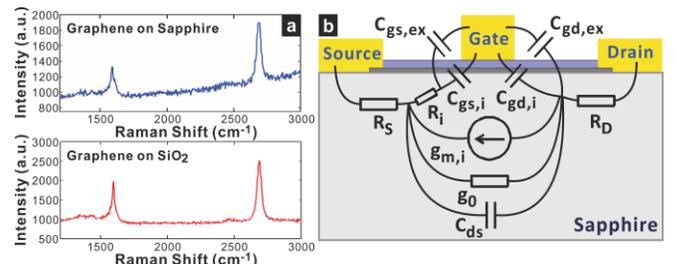

Fig. 1 (a) Raman spectra taken with a 532 nm Nd:YAG laser confirm the presence of single-layer graphene on sapphire. The Raman spectrum of graphene on $SiO_2$ is also shown for comparison. (b) Schematic of the GFET on sapphire with the small-signal equivalent circuit overlaid on top. $R_S$ and $R_D$ are the source and drain access resistances. $R_i$ is the intrinsic resistance. $g_{m,i}$ is the intrinsic transconductance. $g_0 = 1/R_{ds}$ is the output conductance. $C_{ds}$ is the source-drain capacitance. $C_{gs,i}$ and $C_{gd,i}$ are the internal gate-source and gate-drain capacitances. $C_{gs,ex}$ and $C_{gd,ex}$ are the external gate-source and gate-drain capacitances.

This work was partially supported by the ONR GATE MURI program, the MARCO MSD program and the Army Research Laboratory.

Authors are with the Microsystems Technologies Laboratory, Massachusetts Institute of Technology, Cambridge, MA 02139 USA (e-mail: hanw@mtl.mit.edu).

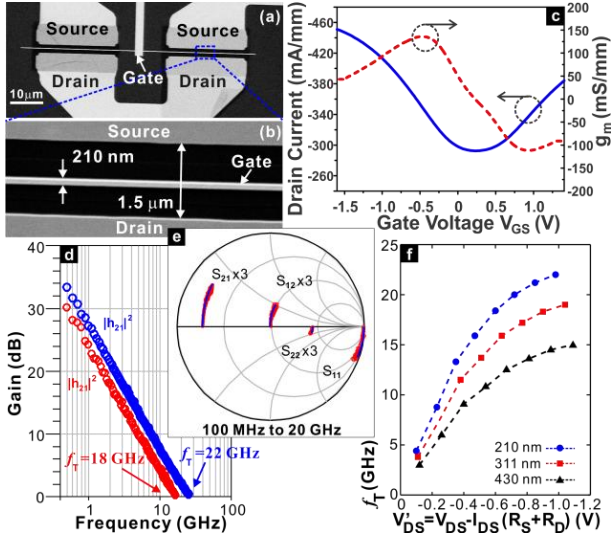

Fig. 2 (a) and (b) SEM images of a device with $L_G$=210 nm and $L_{DS}$=1.5 μm. (c) Transfer characteristics ($I_{DS}$-$V_{GS}$). (d) $f_T$ of this device before (18 GHz, red circles) and after (22 GHz, blue circles) de-embedding the GSG probe pad parasitics. $V_{DS}$=-1.6 V. $V_{GS}$=-0.6 V. (e) The measured S-parameters after de-embedding (red dots) and the S-parameters predicted by the small-signal model (blue curves). (f) Dependence of $f_T$ on drain bias with $V_{GS}$ biased to optimum $g_m$. All measurements were performed in vacuum (1.4×10⁻⁴ Torr).

etching. A gate dielectric consisting of 16 nm $Al_2O_3$ is then formed by naturally oxidizing e-beam evaporated Al. The top gate electrode consists of 60 nm-thick Al. The RF performance of the devices is measured with an N5230A Network Analyzer following short-open-load-through calibration and standard open-short de-embedding process [11]. The device with $L_G$=210 nm shows an $f_T$ of 18 GHz before de-embedding and 22 GHz after de-embedding (Figure 2(d)). The ratio is very close to unity, confirming that the GSG probe pad parasitics are small. Figure 2(e) gives the measured (after de-embedding) and modeled S-parameters, showing excellent agreement. Figure 2(f) shows the dependence of $f_T$ on the intrinsic drain-source bias, $V'_{DS} = V_{DS} - I_{DS}(R_S + R_D)$. Since these GFETs operate in the linear regime, $f_T$ increases with $V'_{DS}$, which leads to the increase in the drain current and hence higher intrinsic transconductance. This behavior is similar to conventional devices in their linear regime [12].

## III. EXTRACTION OF CARRIER TRANSIT DELAY IN GFETS

There are several methods in the literature for extracting carrier transit delays [12-13]. Moll's method [12] is widely used for III-V HEMTs. However, the method is best used for devices operating in the saturation region, in which the dependence of the drain current on the source-drain bias is negligible. The method does not work well with graphene devices because the majority of the GFETs today operate in the linear regime. In addition, the absence of the drain depletion region in graphene transistors makes the concept of drain delay [12] irrelevant to GFETs. The method in [13] requires cold-FET measurement. This is also not suitable for GFETs, which usually have a significant off-state current and never pinch off. Here, we use the method in [14, 15] to extract the carrier transit delay. The $f_T$ of a field-effect transistor is inversely proportional to the total delay ($\tau_{\text{total}}$) of the device, which can be divided into three different components: the intrinsic delay ($\tau_{\text{int}}$), the extrinsic delay ($\tau_{\text{ext}}$), and the parasitic delay ($\tau_{\text{par}}$):

$$\tau_{\text{total}} = \frac{1}{2\pi f_T} = \tau_{\text{int}} + \tau_{\text{ext}} + \tau_{\text{par}} \quad (1)$$

where $\tau_{\text{int}}$ is the time taken by the carrier to cross the intrinsic channel region ($L_G$); $\tau_{\text{ext}}$ is the additional delay associated with the external fringe capacitances and can be interpreted as the additional transit time due to the extended channel region ($\Delta L_G$); and $\tau_{\text{par}}$ is the RC time constant required to charge and discharge the remaining parasitic part of the active device region. The $f_T$ of a device is related to the small-signal circuit parameters as [16]:

$$f_T = \frac{g_{m,i}/(2\cdot\pi)}{[C_{gs}+C_{gd}]\cdot\left[1+\frac{R_S+R_D}{R_{ds}}\right]+C_{gd}\cdot g_{m,i}\cdot(R_S+R_D)} \quad (2)$$

Hence, the three components of the total delay are related to the small-signal circuit parameters (Figure 1(b)) as follows:

$$\tau_{\text{int}} = \frac{C_{gs,i}+C_{gd,i}}{g_{m,i}} \quad (3) \qquad \tau_{\text{ext}} = \frac{C_{gs,ex}+C_{gd,ex}}{g_{m,i}} \quad (4)$$

$$\tau_{\text{par}} = C_{gd}(R_S+R_D)\left[1+\left(1+\frac{C_{gs}}{C_{gd}}\right)\frac{g_o}{g_{m,i}}\right] \quad (5)$$

In this paper, we define the source of hole injection as the source for the GFETs. $C_{gs,i}$ and $C_{gd,i}$ are the internal capacitances. These are the components of $C_{gs}$ and $C_{gd}$ that directly scale with the gate length, while $C_{gs,ex}$ and $C_{gd,ex}$ are the external fringe capacitances, i.e. the components of $C_{gs}$ and $C_{gd}$ that do not change with the gate length. The small-signal capacitances $C_{gs}$ and $C_{gd}$ are first extracted from S-parameters. As shown in Figure 3(a), both the internal and external capacitances of the devices are then extracted from the scaling behavior of $C_{gs}$ and $C_{gd}$ for three GFETs with $L_G$ = 430 nm, 311 nm and 210 nm. The effective lateral electric field in the channel $V'_{DS} = V_{DS} - I_{DS}(R_S + R_D)$ and the intrinsic gate overdrive ($V'_{DS} = -1.0$ V, $V_{GS} = -0.6$ V for the 210 nm device) are kept the same in all three devices in order to achieve similar lateral and vertical electrostatic conditions in the channel for each device. The devices are within a few hundred μm from each other on the same sample and gate lengths are accurately measured by scanning electron microscopy (SEM). $C_{gs}$ and $C_{gd}$ have contributions from both the electrostatic capacitance of the gate dielectric and quantum capacitance of graphene. Unlike conventional devices operating in saturation regime where $C_{gd}$ is much smaller and has a very weak dependence on the gate length

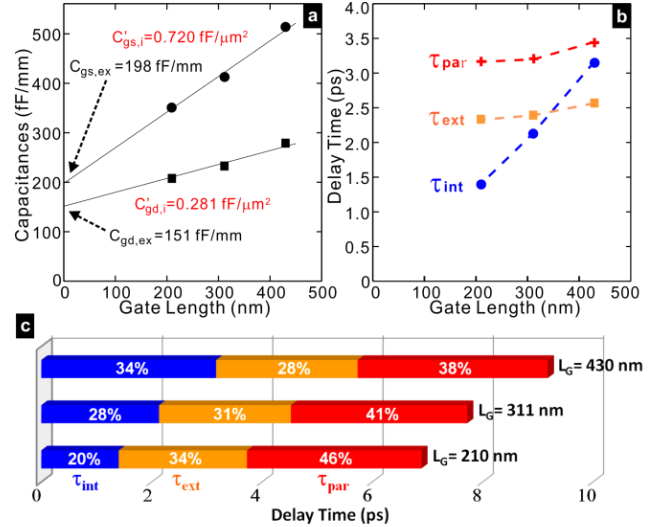

| $L_G$ (nm) | 210 | 311 | 430 |
|---|---|---|---|
| $f_{T,1}=1/2\pi\tau_{\text{int}}$ (GHz) | 107 | 80 | 56 |
| $f_{T,2}=1/(2\pi\tau_{\text{int}}+2\pi\tau_{\text{ext}})$ (GHz) | 40 | 38 | 31 |
| $1/2\pi\tau_{\text{total}}$ (GHz) | 23 | 22 | 19 |
| $f_{T,\text{meas}}$ (GHz) after de-embedding | 22 | 20 | 17 |
| $f_{T,\text{meas}}$ (GHz) before de-embedding | 18 | 17 | 15 |

Fig. 3 (a) Extraction of external gate capacitances ($C_{gs,ex}$, $C_{gd,ex}$) and internal gate capacitances ($C_{gs,i}$, $C_{gd,i}$) from the scaling behavior of $C_{gs}$ and $C_{gd}$. The effective lateral electric field in the channel and the intrinsic gate overdrive $V_{GS} - I_{DS}R_S - V_{G,min}$ are kept the same for each device. (b) Intrinsic ($\tau_{\text{int}}$), extrinsic ($\tau_{\text{ext}}$) and parasitic delays ($\tau_{\text{par}}$) v.s. gate length. (c) Percentage of each delay component in the total delay for each device. (d) frequency performance.

due to minimum charge variation on the drain side in the saturation regime, the majority of the GFETs reported in the literature does not show current saturation and hence have $C_{gd}$ that is a considerable fraction of $C_{gs}$. $C_{gd}$ in GFETs also demonstrates significant dependence on gate length. This Miller capacitance can limit the bandwidth for amplifier applications.

## IV. RESULTS AND DISCUSSIONS

Figure 3(b) shows that the intrinsic delay scales almost linearly with the gate length, as expected, while the extrinsic and parasitic delays both stay relatively constant as gate length changes. Hence, as the gate length reduces, the total delay in these GFETs becomes increasingly dominated by both the extrinsic and parasitic delays while the percentage of the intrinsic delay shrinks (Figure 3(c)). The increasing dominance of the parasitic delay in shorter channel GFETs as shown here agrees with Ref. [17], which shows that the access resistances play a key role in limiting $f_T$ of short channel GFETs. Figure 3(c) shows that the extrinsic delay also becomes more significant in GFETs with shorter channels. Hence, to further improve $f_T$ of RF GFETs, both $\tau_{par}$ and $\tau_{ext}$ need to be reduced. $\tau_{par}$ can be reduced by minimizing the source and drain access resistances, such as using a self-aligned device structure [5]. In addition, both $\tau_{par}$ and $\tau_{ext}$ can be reduced by optimizing the gate thickness and overlap to reduce fringe capacitances.

The intrinsic delay is directly related to the carrier velocity in the channel, which can be evaluated from the slope of the intrinsic delay dependence on $L_G$ in Figure 3(b): $v_h = (\partial \tau_{int}/\partial L_G)^{-1} = 1.24 \times 10^7$ cm/s. While this velocity is extracted in the linear region of sample FETs, it is still much higher than saturation velocity in Si devices [18], demonstrating the great potential of graphene FETs. For a given lateral electric field in the channel, the carrier velocity in linear region is dependent on the carrier mobility. In GFETs, the mobility is mainly limited by the various scattering mechanisms, such as charge impurity scattering, optical phonon scattering, and ripple scattering. Hence, the intrinsic delay for GFETs operating in linear region can be reduced by biasing the channel at a higher lateral electric field to achieve a higher carrier velocity. For GFETs operating at a given bias condition in linear region or for operation in saturation region, the intrinsic delay can be reduced by improving the material quality and by using a better substrate such as boron nitride [19][20] to reduce scattering and improve carrier mobility and carrier velocity.

Figure 3(d) shows the cut-off frequencies for these devices. The measured cut-off frequencies $f_{T,meas}$ after de-embedding agree well with that calculated from the total delay ($1/2\pi\tau_{total}$). $f_{T,2}$ is the cut-off frequency if the access resistances are completely removed. $f_{T,1}$ is directly related to the carrier velocity in the intrinsic channel region and is generally hard to reach in practical devices; but nevertheless, it highlights the great potential of these GFETs. Even with the moderate mobility in the CVD graphene used in this work, $f_{T,1}$ can reach 1 THz if the gate length is reduced to 20 nm. This is a conservative estimate because the carrier transport may become ballistic at such gate length, which can further enhance the frequency performance.

In conclusion, a method for extracting the carrier transit delays in RF transistors is applied to GFETs on sapphire with sub-micrometer gate length. The extraction of intrinsic delay offers a new way to estimate the carrier velocity in the channel. By breaking down the total delay into individual components associated with intrinsic carrier velocity, fringe capacitances, and access region parasitics, this method provides insightful information for device optimization. These three delay components can also serve as figures of merit for comparing the quality of RF GFETs in terms of both the intrinsic material transport property (by using $\tau_{int}$) and the design and quality of the external device structure (by using $\tau_{par}$ and $\tau_{ext}$).

## REFERENCES

[1] K. I. Bolotin, K. J. Sikes, J. Hone, H. L. Stormer, and P. Kim, "Temperature-Dependent Transport in Suspended Graphene," *Physical Review Letters*, vol. 101, no. 9, p. 096802, 2008.
[2] H. Wang, D. Nezich, J. Kong, and T. Palacios "Graphene Frequency Multipliers" *IEEE Electron Device Lett.*, vol. 30, no. 5, May 2009.
[3] H. Wang, et. al., "Graphene-based Ambipolar RF Mixers" *IEEE Electron Device Lett.*, vol. 31, no. 9, Sept. 2010.
[4] Y.-M. Lin, C. Dimitrakopoulos, K. A. Jenkins, D. B. Farmer, H.-Y. Chiu, A. Grill, P. Avouris "100-GHz Transistors from Wafer-Scale Epitaxial Graphene," *Science*, vol. 327, no. 5966, p. 662, Feb. 2010.
[5] L. Liao, Y. Lin, M. Bao, R. Cheng, J. Bai, Y. Liu, Y. Qu, K. Wang, Y. Huang, X. Duan "High-speed graphene transistors with a self-aligned nanowire gate," *Nature*, vol. 467, no. 7313, pp. 305-308, 2010.
[6] J. S. Moon, D. Curtis, M. Hu, D. Wong, C. McGuire, P. M. Campbell, G. Jernigan, J. L. Tedesco, B. VanMil, R. Myers-Ward, C. Eddy, D. K. Gaskill "Epitaxial-Graphene RF Field-Effect Transistors on Si-Face 6H-SiC Substrates," *IEEE Electron Device Letters*, vol. 30, no. 6, 2009.
[7] E. Pallecchi, C. Benz, A. C. Betz, H. v. Löhneysen, B. Plaçais, and R. Danneau, "Graphene microwave transistors on sapphire substrates," *Applied Physics Letters*, vol. 99, p. 113502, 2011.
[8] X. Li, W. Cai, J. An, S. Kim, J. Nah, D. Yang, R. Piner, A. Velamakanni, I. Jung, E. Tutuc, S. K. Banerjee, L. Colombo, R. S. Ruoff, "Large-Area Synthesis of High-Quality and Uniform Graphene Films on Copper Foils," *Science*, vol. 324, no. 5932, Jun. 2009.
[9] A. Aminzadeh, "Excitation Frequency Dependence and Fluorescence in the Raman Spectra of $Al_2O_3$," *Applied Spectroscopy*, vol. 51, no. 6, 1997.
[10] A. Hsu, H. Wang, K. K. Kim, J. Kong, and T. Palacios, "Impact of Graphene Interface Quality on Contact Resistance and RF Device Performance," *IEEE Electron Device Letters*, vol. 32, no. 8, Aug. 2011.
[11] M. C. A. Koolen, J. A. Geelen, and M. P. J. Versleijen, "An improved de-embedding technique for on-wafer high-frequencycharacterization," *Bipolar Circuits and Technology Meeting* 1991, pp. 188-191.
[12] N. Moll, M. R. Hueschen, and A. Fischer-Colbrie, "Pulse-doped AlGaAs/InGaAs pseudomorphic MODFETs," *IEEE Transactions on Electron Devices*, vol. 35, no. 7, pp. 879-886, Jul. 1988.
[13] T. Suemitsu, "An intrinsic delay extraction method for Schottky gate field effect transistors," *IEEE Electron Device Letters*, vol. 25, no. 10, 2004.
[14] J. A. del Alamo and D.-H. Kim, "Is $f_T$ over 1 THz possible?" *Proc. Top. Workshop Heterostruct. Microelectron.*, Nagano, Japan, Aug. 2009.
[15] D. S. Lee, X. Gao, S. Guo, D. Kopp, P. Fay, T. Palacios "300-GHz InAlN/GaN HEMTs With InGaN Back Barrier" *IEEE Electron Device Letters*, vol. 32, no. 11, November 2011.
[16] P. J. Tasker and B. Hughes, "Importance of source and drain resistance to the maximum $f_T$ of millimeter-wave MODFETs," *IEEE Electron Device Letters*, vol. 10, no. 7, pp. 291-293, Jul. 1989.
[17] Y. Q. Wu, Y.-M. Lin, K. A. Jenkins, J. A. Ott, C. Dimitrakopoulos, D. B. Farmer, F. Xia, A. Grill, D. A. Antoniadis, P. Avouris "RF performance of short channel graphene field-effect transistor," *IEDM* 2010, pp. 9.6.1-9.6.3.
[18] A. Lochtefeld and D. A. Antoniadis, "On experimental determination of carrier velocity in deeply scaledNMOS: how close to the thermal limit?" *IEEE Electron Device Letters*, vol. 22, no. 2, pp. 95-97, Feb. 2001.
[19] C. R. Dean, A. F. Young, I. Meric, C. Lee, L. Wang, S. Sorgenfrei, K. Watanabe, T. Taniguchi, P. Kim, K. L. Shepard, and J. Hone "Boron nitride substrates for high-quality graphene electronics," *Nature Nanotechnology*, vol. 5, no. 10, pp. 722-726, Oct. 2010.
[20] H. Wang, et. al., "BN/Graphene/BN Transistors for RF Applications" *IEEE Electron Device Lett.*, vol. 32, no. 9, Sept. 2011.